*Article*

# What is Fair Pay for Executives? An Information Theoretic Analysis of Wage Distributions

## Venkat Venkatasubramanian

Laboratory for Intelligent Process Systems, School of Chemical Engineering, Purdue University, West Lafayette, IN 47907, USA; E-Mail: venkat@ecn.purdue.edu

**Abstract:** The high pay packages of U.S. CEOs have raised serious concerns about what would constitute a fair pay. Since the present economic models do not adequately address this fundamental question, we propose a new theory based on statistical mechanics and information theory. We use the principle of maximum entropy to show that the maximally fair pay distribution is lognormal under ideal conditions. This prediction is in agreement with observed data for the bottom 90%–95% of the working population. The theory estimates that the top 35 U.S. CEOs were overpaid by about 129 times their ideal salaries in 2008. We also provide an insight of entropy as a measure of fairness, which is maximized at equilibrium, in an economic system.

## 1. Introduction

In recent years, there has been great concern over the high pay packages awarded to the Chief Executive Officers (CEOs) of U.S. corporations. The ratio of CEO salary (i.e., total compensation including bonuses and stock options) to that of an average employee has gone up from about 25–40 in the 1970s to as high as 344 in recent years in the U.S. [1]. Compared with minimum wage, the ratio has risen from about 50 in 1965 to about 866 in 2007 [1,2]. However, the ratio has remained around 20–40 in Europe and 10–15 in Japan [3].

All these comparisons, naturally, raise the question of "What is the fair pay for a CEO relative to other employees?" which we try to address in this paper. Clearly, corporate boards and compensation committees seem to think that CEOs deserve such high pay as they are the ones who approve the pay packages. But how does a board arrive at this decision? On what scientific basis does the board make its assessment and decision? Is this a rational, unbiased decision? These concerns lead to an important fundamental question: Is there a rational quantitative framework for evaluating a CEO's value in a corporation?

The usual response to these questions, of course, is that the free market takes care of all this and determines the value of a CEO and the other employees. If an employee feels that his or her pay package is low and unfair, he or she will shop around in the market until he or she receives a fair offer and will move to another organization, subject to constraints such as geographical preferences, career growth opportunities, impact on family, etc. Is the market really efficient in determining these fair values? For markets to be efficient there needs to be free and unfettered flow of voluminous information–information about job openings, compensation packages, growth opportunities, skill sets, etc.–and a very large number of potential candidates who are available to trade jobs.

One can reasonably expect the lower to middle level of the compensation spectrum, e.g., from secretaries to engineers to middle level management professionals, to arrive at a fair wage distribution through an iterative, market-driven, feedback process as there is free and unfettered flow of large amounts of information, and large number of candidates, at these levels. However, it is doubtful that such a distribution is achieved at the very top end of the compensation spectrum as it is a thinly traded market. Further, the details of the pay packages or the process of recruiting candidates are often not very transparent. In their persuasive work, Bebchuk and Fried [4] describe how the "arm's-length contracting" model of executive compensation has broken down, and provide several reasons for this systemic failure. Under such conditions, it should not be surprising that the free market may not function as freely and efficiently as it could at the top end of the pay spectrum. The net result could be over compensation of CEOs and the upper management. But does this happen? If so, how much is this over compensation? Is there a quantitative framework for estimating this?

## 2. Towards a Rational Quantitative Framework for Determining Relative Value of Employees

Now let us examine a couple of compensation scenarios. Consider a computer company that is successful with a number of products and services. All the employees contribute to its overall success in their own ways. The cleaning crew keeps the premises neat, secretarial staff help with organization and communication, engineers develop products and services, marketing and sales personnel bring new orders, accounting and finance department minds the books, management focuses on a winning corporate strategy and execution, and so on. Different people contribute in different ways to the company's overall success. How do we value each one's contribution and reward them suitably? While it is clear that they all contribute, are they all contributing equally? Are some employees contributing more than others? Are some employees more valuable than others? Are some skill sets and experiences more important than others?

In scenario one, let us say that the company's compensation committee has an egalitarian philosophy, values all employees as identically equal, and pays them equally. Everybody in the company, right from the CEO to the cleaning crew member, gets the same salary. While this may sound wonderful in a social justice sense, is this really a fair distribution of the company's profits? In doing so, one makes a very important assumption that every one is contributing *equally* to the overall success of the company. Is this a correct assumption? This assumption implies that the contribution made by the cleaning crew member is exactly equal to that made by the chief architect of the company's successful VLSI chip. This further implies that if the chief architect were to quit tomorrow its effect on the company's prospects would be the same as the cleaning crew member not showing up for work. Obviously, they are not the same. While they both contribute, they ought to be valued differently. Otherwise, disenchanted employees will leave and there will be no company left, despite its noble intentions.

Let us now consider the other extreme. Only the CEO is considered to be most valuable, and everyone else is of minimal value. The CEO gets most of the profits as his or her pay and the rest are paid negligible amounts. Obviously, employees will flee the company in droves and soon there will be no company left.

Clearly, the reality is somewhere in between these two extremes of valuation schemes. But where? What is a *maximally fair* assessment of relative value and the concomitant distribution of profits among the employees?

In the first scenario, even though it may seem like an unbiased assessment, unbiased in the social justice sense (i.e., "all human beings are equal"), it is an *extremely biased* position in the economic sense. From the perspective of economic productivity contributed by different employees, not everyone is equal in importance in their contributions towards the final products and services. Thus, it is unfair to assume that all should be valued equally in their contributions. Similarly, the other case, is also extremely biased to think that only the CEO made everything happen.

Thus, the question of what is a *maximally fair* assessment of relative values of employees reduces to the following: What is the distribution of wages or profits that avoids such biases and assumptions? In other words, what is the *least biased* distribution of attributing value? Since pay packages reflect the perceived values of the employees, one can use pay as a proxy for value. So, the question then is what is the *least biased* distribution of pay?

In a competitive free market environment, companies and employees as rational agents arrive at this distribution iteratively, by a trial-and-error evolutionary feedback process, through the free exchange of information and people between companies and the market environment, until equilibrium is reached. The survival instinct drives people to maximize their values and trade their current jobs for more rewarding ones. Companies also do the same by hiring and firing employees in order to derive better value from them and maximize profits. Therefore, the least biased distribution of relative value is reached in practice via such a market process, *empirically*. That is, the equilibrium distribution is "discovered" through such evolutionary adaptation.

We now ask the following question: instead of the *empirical discovery* approach, can we *design* the least biased distribution *a priori*? That is, can we *predict theoretically* what this distribution *ought* to be at equilibrium? Is there a quantitative framework that could help us address this fundamental question? Such a framework, even if it were limited to ideal conditions, would still be quite valuable as a reference model with respect to which reality may be compared. Such a model would provide us useful quantitative metrics which measure the deviation from ideality. It would help us understand and quantify how *rational and efficient* the market is in discovering this distribution in practice.

At present, there appears to be no such predictive, rational, quantitative framework along the lines discussed above. There is, however, extensive literature on the executive pay packages, income distributions, etc., but typically from the empirical perspective [5-10]. There appears to be no discussion on what a CEO pay package ought to be and why, from a fundamental perspective. Currently, reflecting the popular outrage in the U.S. towards executive compensation, there are suggestions being made to cap the CEO pay at some arbitrary level (such as 25 times the minimum wage [11]) without any rational quantitative analysis justifying why that particular level. Therefore, a rational perspective would be of considerable help in many ways.

It turns out that one can indeed develop such a theory, and somewhat surprisingly, the answer does not come from economics as one might expect, but from the fundamental principles of statistical mechanics and information theory.

## 3. An Information Theoretic Framework

Let us address this question from an information theoretic perspective. Our objective here is not to develop an exact model of the relative value distribution which is impossible at the present as one does not know, and quite possibly may never know, how to measure the value of an individual's contributions precisely. Our primary objective is to formulate a general theory that can provide a rational quantitative framework for a fair assessment of relative values of employees in an organization so that we can *design a priori* the least biased distribution. Our secondary objective is to develop a model that is capable explaining and predicting the character of the relative value distribution as observed in practice.

To get started with such a design perspective, let us examine what we know about this distribution *a priori*. There are many things we do not know about this distribution, but we do have some partial knowledge. Typically, one (or the compensation committee) does know at least four things: (i) total number of employees ($N$) including the CEO, (ii) total amount of money budgeted ($M$) to pay all these employees, (iii) minimum salary ($S_{min}$) received by the lowest employee, often fixed by the minimum wage law, and (iv) the maximum salary ($S_{max}$) cannot exceed $M$. This may not seem like much, but even this partial knowledge can help us a great deal by narrowing the choices, as we shall show, since the value distribution is constrained by this information. In practice, one might have more information but at the very least one would possess this data set.

In order to simplify the analysis, let us consider an ideal situation where one's perceived value $V$ in an organization is captured entirely by his or her salary $S$, i.e., $V = f(S)$. Reality, of course, is much more complicated than this. Titles, awards, peer recognition, perks, etc. can matter a lot as metrics of value in addition to pay. While real life distributions can be distorted by such factors, pay, however, is still the dominant factor in recognizing an employee's perceived value. We assume $S$ to be continuous.

While we claim that pay is a reasonable proxy for perceived value, it is, however*,* not a linear scale. It is quite well known that money has diminishing marginal utility as an incentive and that the value of money has a saturation behavior. This saturation behavior of money has generally been modeled (e.g., as in St Petersburg Paradox [12]) as a logarithmic function. Further, we can also derive this relationship from the observation that since people value money as a proportion, it follows that:

$$V = f(S), \text{ s.t. } dV = dS/f(S)$$

which leads to the perceived value $V_i$ of any employee $i$ as:

$$V_i = C \ln S_i \tag{1}$$

where $S_i$ is his or her *total* annual salary including bonuses and other benefits, and C is a constant of proportionality which we set equal to one. We choose the natural logarithm for mathematical convenience.

Therefore, the central question of "What is the least biased allocation of values among employees?" now reduces to the question of "What is the least biased distribution of *M* dollars among *N* employees as salaries $S_i$?"

Another useful piece of information we have is the range *R* of the values, given by:

$$R \equiv V_{max} - V_{min} \tag{2}$$

$$= \ln S_{max} - \ln S_{min} = \ln M - \ln S_{min} \tag{3}$$

From (3), we can estimate the standard deviation $\sigma$ of the distribution by using the well known Chebychev inequality given by [13-15]:

$$P[-a\sigma < X - \theta < a\sigma] \geq 1 - 1/a^2 \tag{4}$$

where, $\theta$ is the mean of the distribution. By choosing a large enough value for *a* (e.g., $a = 10$), one can estimate $\sigma$ as $R/2a$ with as much confidence as one desires (e.g., for $a = 10$, $P \geq 0.99$). As we shall see, the actual value of *a* is not important for our purposes. Thus, we now know one more thing about this distribution, an estimate of its $\sigma$. Of course, we realize that this is an overestimate since the upper bound *M* for salary is usually not the case. In practice, one would have a much more realistic upper bound. Fortunately, as we shall see, this does not affect our main conclusions.

Now, we can also estimate the mean or the expected value $E[V]$:

$$E[V] = E[\ln S] \tag{5}$$

by using another well known result, namely, Jensen's inequality [14,15], given by:

$$E[\ln S] \leq \ln(E[S]) \tag{6}$$

$$\leq \ln(M/N) \tag{7}$$

which gives us an upper bound for the mean:

$$E[\ln S] = \ln(M/N) \equiv \mu \tag{8}$$

Now that we have gained additional knowledge about this distribution, let us analyze the model derived from these upper bounds. Our central question has thus been reduced to: *"What is the least biased distribution of M dollars among N employees given the constraints $E[lnS] = \mu$ and $E[(lnS)^2] = \sigma^2$ ?"*

This question can be answered by applying the Principle of Maximum Entropy (PME) from information theory [16-18]. Principle of Maximum Entropy states that given some partial information about a random variate *S*, of all the distributions that are consistent with the given information, the *least biased distribution* is the one that has the *maximum entropy* associated with it. Thus, the maximum entropy distribution does not make any unwarranted assumptions or biases about individual values that are not explicitly specified *a priori* as constraints.

Following Shannon's definition of entropy, and applying PME under the above specified constraints using the method of Lagrange multipliers, it is easy to show [19,20] that the least biased distribution is *lognormal*, given by the probability density function:

$$f(S; \mu, \sigma) = \frac{1}{S\sigma\sqrt{2\pi}} e^{-\frac{(\ln S - \mu)^2}{2\sigma^2}} \quad (9)$$

with the mean:

$$E[S] = e^{\mu + \frac{1}{2}\sigma^2} \quad (10)$$

and variance:

$$Var[S] = (e^{\sigma^2} - 1)e^{2\mu + \sigma^2} \quad (11)$$

The entropy *H* of this distribution is given by:

$$H(\mu, \sigma) = \mu + \frac{1}{2}\ln(2\pi e \sigma^2) \quad (12)$$

*Thus, the lognormal distribution of salaries is the least biased, maximally fair, way of distributing pay in an organization under ideal conditions.*

It is also easily seen that the distribution of *lnS* follows a normal distribution with the mean *μ and* variance *σ²*. In other words, the maximally fair distribution *salaries* is lognormal while that of *values* is normal.

We do realize that our estimates of mean and standard deviation are upper bounds and that the true mean *μ\** and standard deviation *σ\** of the population will be lower. Let us assume, for the sake of argument, that through additional information we are provided a better estimate of the mean (= *μ´*) and standard deviation *(= σ´)* which are a lot closer to the true values. Under these conditions, the PME would *still yield a lognormal distribution*, but with a *different* mean (*μ´*) and standard deviation (*σ´*). Thus, the crucial insight here is that even if our guesses about the mean and standard deviation are not very good, the *essential qualitative character of the distribution does not change and remains lognormal* – only the parameters such as the mean and standard deviation change. Further more, it turns out that our initial estimate for the mean as *μ* is not too bad. Since the resulting distribution turned out to be lognormal, the approximation of:

$$E[lnS] = ln[E(S)] \quad (13)$$

is actually quite good.

How does this prediction compare with reality? The pay or wage distribution data for individual companies are generally not available and are often regarded as proprietary information. However, one can compare with aggregate data reported in the literature gathered from income tax filings for millions of people. For large corporations that employ tens of thousands of people (e.g., Fortune 500 companies), it is reasonable to expect the essential character of the pay distribution for such corporations to be quite close to that of the population data. The mean and variance might differ but it is reasonable to expect the essential qualitative nature to remain the same.

Our prediction of a lognormal distribution is in good agreement with the population data [5-10, 24] for the bottom part of the spectrum. It's been reported that typically the bottom 90%–95% of the income distribution follows the lognormal distribution while the top 5%–10% follows a Pareto, or power law, distribution. As noted earlier, these are "discovered" empirically in practice through the iterative free market-driven process. Again, we expect the market to be functioning freely and efficiently for the lower and the middle parts of the distribution. Hence, it is not surprising that the maximally fair distribution is enforced by the market forces for those employees. However, it is still quite remarkable that our prediction fits the observed data for a very high percentage (~90%–95%) of the population, even though our model was developed under idealized assumptions thereby potentially limiting its real-world applicability. It is also very encouraging to see that in practice a great majority of the employees are being treated fairly.

Regarding the top 5%–10%, the situation is more complicated as the aggregate income data is confounded by a combination of both wage and investment incomes [21-25]. One needs wage only data from companies to test the model for this end of the spectrum conclusively. At this juncture, it is important to differentiate between the results reported in the literature on income distributions from wealth distributions. Even though both seem to have similar structures, namely, a lognormal body with a Pareto tail [21-25], our focus in this paper is on wage distribution. We believe the proposed framework could be applied to the study of wealth distributions as well since one's wage income is the first step towards accumulating wealth, and therefore, patterns in wage income generation could lead to similar patterns in wealth creation. However, since wealth creation also includes returns (including appreciation) from investments, the underlying economic mechanisms for wealth creation are different from wage distribution ones.

Others have proposed thermodynamically inspired models for the emergence of income and wealth distributions [21-33]. Our contribution, however, takes a different perspective, namely, an information theoretic one, and asks a very different question, namely, "What is the maximally fair distribution of wages among employees?", even though our methodology, too, draws on concepts from thermodynamics and statistical mechanics.

As others have shown [21-31], the bottom 90%–95% of the income data seem to fit both lognormal and Gibbs-Boltzmann distributions. Economists seem to favor the lognormal distribution while econophysicists prefer the Gibbs-Boltzmann distribution [24,27]. Our theory suggests that the lognormal distribution is perhaps the right candidate and offers justification based on fundamental principles which was lacking in the earlier studies. Furthermore, it captures the essential concepts of economics such as value, fairness, and marginal utility of money, which are missing in the purely thermodynamic formalisms that have been proposed before [26,32,33].

## 4. What is Fair Pay for CEOs?

This analysis has important implications for designing fair pay packages for the top management (top 5%–10%) including the CEO. Under the ideal conditions discussed above, the top management's pay also should fall in line with the rest of the employees wages on a lognormal distribution. Otherwise, we are treating the rest of the company unfairly by over paying for the services of the top management. Further, the company is also wasting substantial resources by over compensating the senior management [4]. Then there is also the risk of creating poor morale among the vast majority of the employees [4,34].

So, how would a compensation committee go about determining a fair pay structure for the top management? The obvious procedure would be is to best fit the salary data to a lognormal distribution for the bottom 90%–95% of the employees and determine its mean and variance. This is the distribution that the rest of the data (i.e., top 5%–10%) ought to follow, so that the entire population of employees is treated fairly on the same basis. The CEO's salary can be determined directly from this extrapolated lognormal distribution plot or by calculating the standard deviation for the CEO's data position ($\sigma_{CEO}$) on the bell curve and computing the corresponding salary for it. Adjustments to this ideal pay package can be made as needed, to account for individual talents and accomplishments, by relating them to both short term and long term performance targets.

Since the salary distribution data for companies are not available, we present as an example (Tables 1 and 2) a number of plausible salary scenarios, and estimate what a fair pay for a CEO ought to be for the top CEOs whose 2008 pay packages were published by the New York Times recently [35]. We consider 35 companies from the top 50 for which reliable employee data were available from various websites. Since *lnS* follows a normal distribution, we perform our calculations using the standard normal distribution tables which make the calculations easier. In our calculations, we assume that:

$$ln\ (mean\ salary) - ln\ (minimum\ salary) = 3\sigma \tag{14}$$

From this we estimate $\sigma$ and calculate the CEO salary by computing the corresponding $\sigma_{CEO}$ upper bound for which the outside area under the bell curve (i.e. under the far right tail) equals 1/N, where N is the total number of employees in the company. For example, for Motorola which has about 64,000 employees, 1/64000 = 0.0000156 corresponds to a position on the bell curve that is 4.16 standard deviation (i.e., $\sigma_{CEO} = 4.16\sigma$) away to the right of the mean, thereby fixing the ideal CEO salary. We *emphasize* that these are only rough estimates, as we lack detailed salary distribution data which are needed to estimate the $\mu$, $\sigma$, and $\sigma_{CEO}$ more accurately. Once we have the actual pay distribution, one can apply much more rigorous statistical methodologies to better estimate these parameters and the CEO's pay.

Nevertheless, even these rough estimates are quite illuminating. We consider two cases: (i) in Table 1, the minimum salary is taken to be the minimum wage of $6.55/hr or $13,100/year assuming a 2,000 work hours/yr basis; (ii) in Table 2, minimum salary is set equal to $25,000/yr, which is perhaps more typical in many corporations. For each of these cases, we consider several mean annual salary scenarios: $40K, $60K, $80K, and $100K. The CEO pay ratios are computed by dividing the actual, and the ideal, annual CEO pay by the minimum annual salary.

**Table 1.** Comparing top CEO salaries with minimum wage: actual vs ideal ratios.

| Company | CEO's Total Pay Incl salary+bonus in $millions 2008 | Total Employees (N) estimates from various websites 2008 | Outside Area (1/N) from Normal Distrn Z-tables | CEO Sigma Estimate From Z-Tables | Actual CEO Pay Ratio | Minimum salary = $13100 | | | |
|---|---|---|---|---|---|---|---|---|---|
| | | | | | | Ideal CEO Pay Ratio Mean Salary=$40K Scenario 1 | Ideal CEO Pay Ratio Mean Salary=$60K Scenario 2 | Ideal CEO Pay Ratio Mean Salary=$80K Scenario 3 | Ideal CEO Pay Ratio Mean Salary=$100K Scenario 4 |
| Motorola | 104.4 | 64000 | 0.00001563 | 4.16 | 7969 | 14.3 | 37.7 | 74.9 | 127.5 |
| Oracle | 84.6 | 50000 | 0.00002000 | 4.11 | 6458 | 14.1 | 36.7 | 72.6 | 123.1 |
| Walt Disney | 51.1 | 150000 | 0.00000667 | 4.35 | 3901 | 15.4 | 41.6 | 84.2 | 145.4 |
| American Express | 42.8 | 30162 | 0.00003315 | 3.99 | 3267 | 13.5 | 34.6 | 67.6 | 113.7 |
| Citi Group | 38.2 | 324850 | 0.00000308 | 4.52 | 2916 | 16.4 | 45.2 | 93.0 | 162.7 |
| Hewlett Packard | 34.0 | 321000 | 0.00000312 | 4.52 | 2595 | 16.4 | 45.2 | 93.0 | 162.7 |
| News Corp | 30.1 | 53000 | 0.00001887 | 4.12 | 2298 | 14.1 | 36.9 | 73.1 | 124.1 |
| Honeywell | 28.7 | 128000 | 0.00000781 | 4.32 | 2191 | 15.2 | 40.9 | 82.5 | 142.1 |
| Proctor & Gamble | 25.6 | 138000 | 0.00000725 | 4.34 | 1954 | 15.3 | 41.3 | 83.5 | 144.0 |
| Abbott | 25.1 | 68697 | 0.00001456 | 4.18 | 1916 | 14.4 | 38.1 | 75.8 | 129.3 |
| Lockheed Martin | 22.9 | 146000 | 0.00000685 | 4.35 | 1748 | 15.4 | 41.5 | 84.0 | 145.0 |
| eBay | 22.5 | 7769 | 0.00012872 | 3.65 | 1718 | 11.9 | 29.1 | 55.1 | 90.3 |
| Anadarko Petroleum | 22.2 | 4000 | 0.00025000 | 3.48 | 1695 | 11.1 | 26.7 | 49.7 | 80.5 |
| United Technologies | 22.0 | 223100 | 0.00000448 | 4.44 | 1679 | 15.9 | 43.5 | 88.6 | 154.1 |
| Bristol Myers Squibb | 21.8 | 42000 | 0.00002381 | 4.07 | 1664 | 13.9 | 36.0 | 70.9 | 120.0 |
| Hess | 21.3 | 13300 | 0.00007519 | 3.79 | 1626 | 12.5 | 31.3 | 59.9 | 99.3 |
| Johnson & Johnson | 21.1 | 118700 | 0.00000842 | 4.30 | 1611 | 15.1 | 40.5 | 81.5 | 140.2 |
| IBM | 21.0 | 398455 | 0.00000251 | 4.57 | 1603 | 16.7 | 46.4 | 95.9 | 168.3 |
| Verizon | 19.9 | 234971 | 0.00000426 | 4.45 | 1519 | 16.0 | 43.7 | 89.2 | 155.2 |
| Coca Cola | 19.6 | 90500 | 0.00001105 | 4.24 | 1496 | 14.8 | 39.3 | 78.6 | 134.6 |
| Avon | 19.5 | 42000 | 0.00002381 | 4.07 | 1489 | 13.9 | 36.0 | 70.9 | 120.0 |
| Cisco | 18.8 | 32160 | 0.00003109 | 4.01 | 1435 | 13.6 | 34.9 | 68.4 | 115.2 |
| Qualcomm | 18.6 | 11932 | 0.00008381 | 3.76 | 1420 | 12.4 | 30.8 | 58.8 | 97.3 |
| General Dynamics | 18.0 | 83500 | 0.00001198 | 4.22 | 1374 | 14.7 | 38.9 | 77.6 | 132.8 |
| CVS | 17.4 | 160000 | 0.00000625 | 4.37 | 1328 | 15.5 | 41.9 | 85.0 | 147.0 |
| Merck | 17.3 | 58900 | 0.00001698 | 4.15 | 1321 | 14.3 | 37.5 | 74.4 | 126.7 |
| Prudential | 16.3 | 49616 | 0.00002015 | 4.11 | 1244 | 14.1 | 36.8 | 72.7 | 123.3 |
| Deere | 16.2 | 52022 | 0.00001922 | 4.12 | 1237 | 14.1 | 36.9 | 73.1 | 124.1 |
| AT&T | 15.0 | 302660 | 0.00000330 | 4.51 | 1145 | 16.3 | 45.0 | 92.5 | 161.6 |
| ADM | 15.0 | 27600 | 0.00003623 | 3.97 | 1145 | 13.4 | 34.2 | 66.8 | 112.1 |
| Pepsi | 14.9 | 198000 | 0.00000505 | 4.41 | 1137 | 15.7 | 42.8 | 87.1 | 151.0 |
| Johnson Comtrols | 14.9 | 140000 | 0.00000714 | 4.34 | 1137 | 15.3 | 41.3 | 83.5 | 144.0 |
| Pfizer | 14.8 | 86600 | 0.00001155 | 4.22 | 1130 | 14.7 | 38.9 | 77.6 | 132.8 |
| Boeing | 14.8 | 162200 | 0.00000617 | 4.37 | 1130 | 15.5 | 41.9 | 85.0 | 147.0 |
| Burlington | 14.6 | 40000 | 0.00002500 | 4.06 | 1115 | 13.8 | 35.8 | 70.5 | 119.2 |
| Average | 26.4 | | | | 2017 | 14.6 | 38.6 | 77.1 | 131.9 |
| Berkshire | 0.2 | 246083 | 0.00000406 | 4.46 | 15 | 16.0 | 43.9 | 89.7 | 156.2 |

Even in the most optimistic scenario (Table 1: Minimum wage Scenario #4), we see that these CEOs have been overpaid considerably. In Table 2, Scenario #2, which is perhaps more close to reality, the excesses are very large. On an average, the actual CEO pay ratio is 1,057 while the ideal value is about 8.2 – i.e., about 129 times the ideal value. In 2008, S&P 500 CEOs averaged about $10.0 million, with an average CEO pay ratio of about 400 (on the $25,000/year minimum salary basis) which is about 50 times the ideal value.

**Table 2.** Comparing top CEO salaries with minimum salary of $25 k/yr: actual vs. ideal ratios.

| Company | CEO's Total Pay Incl salary+bonus in $millions 2008 | Total Employees (N) estimates from various websites 2008 | Outside Area (1/N) from Normal Distrn Z-tables | CEO Sigma Estimate From Z-Tables | Actual CEO Pay Ratio | Minimum salary = $25000 | | | |
|---|---|---|---|---|---|---|---|---|---|
| | | | | | | Ideal CEO Pay Ratio Mean Salary=$40K Scenario 1 | Ideal CEO Pay Ratio Mean Salary=$60K Scenario 2 | Ideal CEO Pay Ratio Mean Salary=$80K Scenario 3 | Ideal CEO Pay Ratio Mean Salary=$100K Scenario 4 |
| Motorola | 104.4 | 64000 | 0.00001563 | 4.16 | 4176 | 3.1 | 8.1 | 16.0 | 27.3 |
| Oracle | 84.6 | 50000 | 0.00002000 | 4.11 | 3384 | 3.0 | 7.9 | 15.7 | 26.6 |
| Walt Disney | 51.1 | 150000 | 0.00000667 | 4.35 | 2044 | 3.2 | 8.5 | 17.3 | 29.9 |
| American Express | 42.8 | 30162 | 0.00003315 | 3.99 | 1712 | 3.0 | 7.7 | 15.0 | 25.2 |
| Citi Group | 38.2 | 324850 | 0.00000308 | 4.52 | 1528 | 3.2 | 9.0 | 18.4 | 32.2 |
| Hewlett Packard | 34.0 | 321000 | 0.00000312 | 4.52 | 1360 | 3.2 | 9.0 | 18.4 | 32.2 |
| News Corp | 30.1 | 53000 | 0.00001887 | 4.12 | 1204 | 3.0 | 8.0 | 15.8 | 26.8 |
| Honeywell | 28.7 | 128000 | 0.00000781 | 4.32 | 1148 | 3.1 | 8.5 | 17.1 | 29.4 |
| Proctor & Gamble | 25.6 | 138000 | 0.00000725 | 4.34 | 1024 | 3.2 | 8.5 | 17.2 | 29.7 |
| Abbott | 25.1 | 68697 | 0.00001456 | 4.18 | 1004 | 3.1 | 8.1 | 16.2 | 27.5 |
| Lockheed Martin | 22.9 | 146000 | 0.00000685 | 4.35 | 916 | 3.2 | 8.5 | 17.3 | 29.8 |
| eBay | 22.5 | 7769 | 0.00012872 | 3.65 | 900 | 2.8 | 7.0 | 13.2 | 21.6 |
| Anadarko Petroleum | 22.2 | 4000 | 0.00025000 | 3.48 | 888 | 2.8 | 6.6 | 12.3 | 19.9 |
| United Technologies | 22.0 | 223100 | 0.00000448 | 4.44 | 880 | 3.2 | 8.8 | 17.9 | 31.1 |
| Bristol Myers Squibb | 21.8 | 42000 | 0.00002381 | 4.07 | 872 | 3.0 | 7.9 | 15.5 | 26.2 |
| Hess | 21.3 | 13300 | 0.00007519 | 3.79 | 852 | 2.9 | 7.2 | 13.9 | 23.0 |
| Johnson & Johnson | 21.1 | 118700 | 0.00000842 | 4.30 | 844 | 3.1 | 8.4 | 16.9 | 29.1 |
| IBM | 21.0 | 398455 | 0.00000251 | 4.57 | 840 | 3.3 | 9.1 | 18.8 | 33.0 |
| Verizon | 19.9 | 234971 | 0.00000426 | 4.45 | 796 | 3.2 | 8.8 | 17.9 | 31.2 |
| Coca Cola | 19.6 | 90500 | 0.00001105 | 4.24 | 784 | 3.1 | 8.3 | 16.5 | 28.3 |
| Avon | 19.5 | 42000 | 0.00002381 | 4.07 | 780 | 3.0 | 7.9 | 15.5 | 26.2 |
| Cisco | 18.8 | 32160 | 0.00003109 | 4.01 | 752 | 3.0 | 7.7 | 15.1 | 25.5 |
| Qualcomm | 18.6 | 11932 | 0.00008381 | 3.76 | 744 | 2.9 | 7.2 | 13.7 | 22.7 |
| General Dynamics | 18.0 | 83500 | 0.00001198 | 4.22 | 720 | 3.1 | 8.2 | 16.4 | 28.1 |
| CVS | 17.4 | 160000 | 0.00000625 | 4.37 | 696 | 3.2 | 8.6 | 17.4 | 30.1 |
| Merck | 17.3 | 58900 | 0.00001698 | 4.15 | 692 | 3.1 | 8.0 | 16.0 | 27.2 |
| Prudential | 16.3 | 49616 | 0.00002015 | 4.11 | 652 | 3.0 | 8.0 | 15.7 | 26.7 |
| Deere | 16.2 | 52022 | 0.00001922 | 4.12 | 648 | 3.0 | 8.0 | 15.8 | 26.8 |
| AT&T | 15.0 | 302660 | 0.00000330 | 4.51 | 600 | 3.2 | 8.9 | 18.4 | 32.1 |
| ADM | 15.0 | 27600 | 0.00003623 | 3.97 | 600 | 3.0 | 7.6 | 14.9 | 25.0 |
| Pepsi | 14.9 | 198000 | 0.00000505 | 4.41 | 596 | 3.2 | 8.7 | 17.7 | 30.6 |
| Johnson Comtrols | 14.9 | 140000 | 0.00000714 | 4.34 | 596 | 3.2 | 8.5 | 17.2 | 29.7 |
| Pfizer | 14.8 | 86600 | 0.00001155 | 4.22 | 592 | 3.1 | 8.2 | 16.4 | 28.1 |
| Boeing | 14.8 | 162200 | 0.00000617 | 4.37 | 592 | 3.2 | 8.6 | 17.4 | 30.1 |
| Burlington | 14.6 | 40000 | 0.00002500 | 4.06 | 584 | 3.0 | 7.8 | 15.4 | 26.1 |
| Average | 26.4 | | | | 1057 | 3.1 | 8.2 | 16.3 | 27.9 |
| Berkshire | 0.2 | 246083 | 0.00000406 | 4.46 | 8 | 3.2 | 8.8 | 18.0 | 31.3 |

Obviously, we do expect real-world economic systems to deviate from ideal systems, thus necessitating larger pay packages for CEOs. But would they deviate so much that the actual CEO pay ratio is 50 to 129 times the ideal benchmark? That's hard to believe particularly when for the rest of the employees, i.e., for the bottom 90%–95%, their compensation follows the ideal lognormal distribution. It appears that the market, which seems to function quite fairly and efficiently for the bottom 90%–95% of the employees, does not appear to work as well for the top end of the spectrum. We now have a rational scientific basis for supporting, in fact quantifying, what people have argued for quite sometime about executive pay excesses and the breakdown of the efficient market at the top management levels. However, further studies along the lines suggested below are needed to understand this non-ideal behavior in greater depth. We need to be able to determine, using a rational analytic framework, how much excess pay over the ideal limit should be paid to the executives as a function of the non-idealities.

As a contrast, it is quite interesting to note that Mr. Warren Buffet, the CEO of Berkshire Hathaway and an outspoken critic of executive pay excesses, drew a salary of $200K in 2008. This makes his pay ratio 8:1 (Table 2), which fits the ideal benchmark estimate (Scenario 2) almost exactly.

It is also instructive to note that, in 2006, according to *The Wall Street Journal* [36], the average CEO pay ratio was about 11:1 in Japan, 15:1 in France, 20:1 in Canada, and 22:1 in Britain, which are not that far off (compared to the U.S. ratios) from the ideal benchmark estimates. As noted in the introduction, even in the U.S. the CEO pay ratios were much more reasonable and in general agreement with the ideal values, in 1960s and early 1970s. Thus, the executive pay excesses appear to be a recent phenomenon in the U.S., perhaps due to the reasons argued by Bebchuk and Fried. This appears to be another valuation bubble – the CEO valuation bubble, much like the ones we have witnessed in stocks, real estate, commodities, etc. While the emphasis of this section has been on CEO pay, the observations made here are applicable for the entire senior management in a corporation.

## 5. Conclusions and Recommendations

For about two decades or so, it's been observed that the CEOs of U.S. companies are being paid rather large pay packages. It seems unfair but is it really unfair? How can one tell? On what basis? What, then, would be a fair salary and why?

We could not answer such questions before, as we lacked a rational quantitative framework. The proposed information theoretic framework helps us address these questions and predicts the ideal wage distribution to be lognormal, which agrees well with observed data for the bottom 90%–95% of the pay distribution. Like the minimum wage law which sets a lower limit, our theory suggests what the maximum wage ought to be under ideal conditions. Of course, there are always exceptional individuals who may deserve a higher reward. This theory provides a rational basis for *setting the fair base pay scales for the top management* (indeed for everyone in the organization) and any added incentive pay package might then be linked to measureable and meaningful performance metrics that promote long term survival and growth of the organization. This result also addresses some thorny issues in pay compression [34] by identifying the optimal pay compression policy as a lognormal distribution. However, further studies are needed to understand these issues much better.

This theory is not valid for small, highly entrepreneurial, organizations where a handful of employees (e.g., the founders of a start-up) are demonstrably much more valuable than the others. But this is not the case for many large organizations (e.g., Fortune 500 companies) with tens to hundreds of thousands of employees, where the CEO is often another hired hand.

Though developed under certain simplifying assumptions, the proposed theory could still be useful as a framework for designing tax policy, corporate governance, and public policy guidelines. It can be used as a starting point to design the ideal, maximally fair, distribution which then can be further adapted and tuned to meet real-world constraints. Companies may even use this as a recruiting and retention tool to show how committed they are to a fair treatment of all employees. Shareholders may require companies to publish this information in their annual reports and could use it to demand justification for any excess pay paid to the top management.

As noted, simplifying assumptions were made since our objective was to develop a general theoretical framework and identify general principles that are not restricted by domain specific details and constraints. Clearly, the next steps are to conduct more comprehensive studies of salary distributions in various organizations in order to understand in greater detail the deviations from ideality in the market place. Agencies such as the Bureau of Labor Statistics and National Bureau of Economic Research could organize task forces to gather wage data from various companies and organizations. The data should be so grouped to analyze wage distribution patterns across several dimensions such as: (i) organization size—small, medium, large, and very large number of employees, (ii) different industrial sectors, (iii) different types such as private corporations, governments (state and federal), non-profit organizations, etc. Similar studies should be conducted in other countries as well so that we can better understand global patterns. We would expect the ideal conditions proposed by this theory to be better satisfied in non-profit organizations such as universities, charitable foundations, and governmental agencies. The applicability of this theory is not limited to pay distribution alone. It is also applicable to other situations (e.g. optimal resource allocation) that arise in economics and social sciences regarding equality and fairness.

Our contribution is mainly a conceptual one, resulting from the generalization of the laws of statistical thermodynamics to teleological systems, such as economic systems, proposed by this author [37]. In this framework, which we call *statistical teleodynamics*, the first law of thermodynamics is generalized to an information theoretic interpretation to yield performance constraints, such as *$E(lnS) = \mu$ and $E[(lnS)^2] = \sigma^2$,* which a teleological system must satisfy. As for the second law, Shannon and Jaynes liberated entropy from its narrow thermodynamic view to a broader information theoretic interpretation as a measure of uncertainty. In our contribution, we propose the use of entropy as a measure of *fairness* [38] in economic systems. Thus, maximizing entropy is the same as maximizing fairness collectively in economic systems. Economic equilibrium is reached under this condition when the participants feel they have received a maximally fair deal given the constraints. As we all know, fairness is a fundamental economic principle that lies at the foundation of the free and efficient market system. It is so vital to the proper functioning of the markets that we have regulations and watchdog agencies that breakup and punish unfair practices such as monopolies, collusion, and insider trading. Thus, it is eminently reasonable, indeed reassuring, to find that maximizing fairness, i.e. maximizing entropy, is the condition for achieving economic equilibrium.

In the past, there have been many attempts to find a suitable interpretation of entropy for economic systems without much success [39,40]. In these attempts, one typically wrote down equations in economics that mirrored and mapped expressions in thermodynamics for entropy, energy, temperature, etc.—but no identification of entropy in terms of meaningful economic concepts was made. Just as entropy is a measure of disorder in thermodynamics and uncertainty in information theory, what does entropy mean in economics? Neither interpretation, disorder nor uncertainty, makes much sense in the economic context. Economic systems work best when they have orderly markets. Why then would anyone want to maximize disorder? Similarly, economic systems work best when there is less uncertainty. Why then would anyone want to maximize uncertainty? The inability to address this crucial issue has been a major conceptual hurdle for decades in making real progress along this line of enquiry.

We believe that identifying entropy as a measure of fairness, which is a fundamental economic principle, is a significant advancement made by our theory in bridging the gap and showing the intimate connection between statistical thermodynamics, information theory, and economics.

Intuitively, it is now clear that any deviation from the equilibrium wage distribution would result in less of a fair deal for the participants overall, and therefore, is not likely to happen—e.g., imagine all employees voluntarily accepting lower salaries even though they are offered higher salaries, which is unlikely to happen spontaneously. Conceptually, this is similar to molecules in an isolated system collectively retreating to and staying in a *small sub-space* of the *accessible phase space spontaneously*, which, of course, is negated by the second law of thermodynamics.

It is instructive to compare our theory of fairness in economic systems with the theories of John Rawls [42] and Robert Nozick [43] on equality, fairness and justice. Rawls arrives at his first principle of equality among people for basic liberties through the application of his 'veil of ignorance' concept. This he views as a risk averse decision taken by a group of rational agents in the 'original position'. In our theory we arrive at the same result by applying the principle of maximum entropy. The 'veil of ignorance', i.e., a veil of maximum uncertainty, is essentially the principle of maximum entropy. Since maximizing entropy is the same as maximizing fairness, Rawls' first principle of equality is derived as the maximally fair assignment and not as a risk averse outcome. Thus, the principle of equality emerges out of fairness, not out of fear, in our theory.

Regarding Rawls' second principle, we first focus on the 'difference principle' part, which is concerned with the distribution of wealth and income. In this, Rawls stipulates that a just society is one wherein the social and economic inequalities are to be arranged such that they are to be of the greatest benefit to the least-advantaged members of society, thereby preferring them more over the others. As Nozick and others have argued, this treats the more-advantaged members of the society in an unfair manner, thereby violating the equality principle. In our theory, by maximizing fairness all are treated equally, subject to the given constraints. Finally, Rawls' requirement that "all positions and offices are open to all" is equivalent to the requirement that all parts of the *accessible phase space* are available to the agents equally in our theory.

Thus, the application of maximum fairness principle results in consistent outcomes of equality and fairness in both cases. The Rawlsian framework is equivalent to our statistical teleodynamics theory in the economic context with the important difference that all conflicts are resolved using the maximum fairness principle. Further, the former is a subset of the latter as Rawls does not address constraints imposed by a market environment or the condition for attaining an equilibrium distribution.

There are other key differences as well which show up in the economic inequality context. First of all, in both Rawlsian and Nozickian approaches, there is no quantification of the concepts of fairness, equality, or justice. They are treated in a qualitative manner. In our theory, we identify fairness as entropy which can be quantified and calculated for a given wage distribution. Secondly, in our theory we show how the maximally fair distribution can be determined given certain information or constraints. Since our theory makes testable predictions about what the maximally fair wage distribution ought to be in a free market environment, it can be used not only to verify the prediction but also to test whether the free market is indeed behaving freely. Such analytical predictions are absent in the theories of Rawls and Nozick, making them harder to verify in real-life markets. Finally, our theory shows how these seemingly disparate systems of a large number of interacting entities, namely, thermodynamic, information, and economic systems (and perhaps even societal systems), are all governed by the same general concepts and principles in a unified analytic framework.

Though our theory was inspired [37] by concepts from statistical thermodynamics and information theory, it is rather remarkable that the resultant statistical teleodynamic framework shares a close resemblance with the structure and elements of the theories of Rawls and Nozick. Indeed, the present author developed his theory unaware of the contributions of Rawls, Nozick, and others in political philosophy, until it was pointed out by one of the reviewers of this paper. In the context of economics (and may be in the broader context of sociology as well, but this has not been developed here as it is not the objective of this paper), we believe that the statistical teleodynamic framework naturally combines the central ideas of Rawls and Nozick, with the exception that the difference principle favoring the weak is discarded. Instead, we have the principle of maximum fairness, i.e., the maximum entropy principle, which too can be seen as a "difference principle" except that it treats every one fairly without favoring any particular group. This is the viewpoint advocated by Nozick. Thus, our theory naturally combines these two approaches even though its development was not motivated by this objective.

Our analysis shows that a certain amount of seeming inequality of pay is inevitable in organizations. Given this reality, the lognormal distribution is the fairest inequality of pay. One may view our result as an "economic law" in the statistical thermodynamics sense. The free market will 'discover' and obey this economic law if allowed to function freely and efficiently without collusion like practices or other such unfair interferences. This result is the economic equivalent to the Boltzmann distribution of the energy landscape for ideal gases. In spirit, it's like the Boyle's law for ideal gases which ignores factors such as intermolecular forces, molecular size, etc., but nevertheless provides a useful basis for developing models for non-ideal systems. In a similar manner, our theory has its obvious limitations and does not take in to account industry or company specific factors, complexities of human interactions, competition and other market conditions, and so on. However, we present it with the hope of stimulating further research to examine its implications in greater depth and breadth for a wide variety of contexts in economics and social sciences.

## Acknowledgments

The author is grateful to William Masters, Wallace Tyner, and the reviewers for their valuable comments on the manuscript. He is also thankful to Sujay Sanghavi and Peter Caithamer for clarifications on Jensen's inequality. This contribution is dedicated, with profound gratitude, to Prof. Edwin T. Jaynes *in memoriam*, whose pioneering research in probability theory, statistical mechanics, and information theory made this work possible.

## References

1. Anderson, S.; Cavanagh, J.; Collins, C.; Pizzigati, S.; Lapham, M. *Executive Excess*; Institute for Policy Studies: Washington, DC, USA, 2008.
2. Mishel, L. *State of Working America*; Cornell University Press: Ithaca, NY, USA, 2005.

*Entropy* **2009**, *11* **780**3. Kato, T. *CEO Compensation and Firm Performance in Japan*; Columbia Business School: New York, NY, USA, April 2003.
4. Bebchuk, L.A.; Fried, J.M. Pay without performance: overview of the issues. *J. App. Corp. Finan.* **2005**, *17*, 8-23.
5. Champernowne, D.G. A model of income distribution. *Econ. J.* **1953**, *63*, 318-351.
6. Champernowne, D.G.; Cowell, F.A. *Economic Inequality and Income Distribution*; Cambridge University Press: Cambridge, UK, 1999.
7. Pareto, V. *Cours d'economie Politique*; F. Rouge: Lausanne, Switzerland, 1897.
8. Gini, C. Measurement of inequality and incomes. *Econ. J.* **1921**, *31*, 124-126.
9. Montroll, E.W.; Shlesinger, M.F. On 1/f noise and other distributions with long tails. *Proc. Natl. Acad. Sci. USA* **1982**, *79*, 3380-3383.
10. *Econophysics of Wealth Distributions*. Chatterjee, A., Yarlagadda, S., Chakrabarti, B.K., Eds.; Springer Verlag: Milan, Italy, 2005.
11. *Income Equity Act (H.R. 3876)*. Anderson, S., Cavanagh, J., Collins, C., Pizzigati, S., Lapham, M., Eds.; Institute for Policy Studies: Washington, DC, USA, 2008.
12. Weirich, P. The St. Petersburg gamble and risk. *Theor. Decis.* **1984**, *17*, 193-202.
13. Hozo, S.P.; Djulbegovic, B.; Hozo, I. Estimating the mean and variance from the median, range, and the size of a sample. *BMC Medical Research Methodology* **2005**, *5*, 1-10.
14. Hogg, R.V.; Craig, A.T. *Introduction to Mathematical Statistics*, 5th ed.; Macmillan: New York, NY, USA, 1995.
15. Mood, A.M.F.; Graybill, F.A.; Boes, D.C. *Introduction to the Theory of Statistics*, 3rd ed.; McGraw-Hill: New York, NY, USA, 1974.
16. Jaynes, E.T. Information theory and statistical mechanics. *Phys. Rev.* **1957**, *106*, 620-630.
17. Jaynes, E.T. Information theory and statistical mechanics, II. *Phys. Rev.* **1957**, *108*, 171-190.
18. Shannon, C.E.; Weaver, W. *Mathematical Theory of Communication*; University of Illinois Press: Urbana, IL, USA, 1949.
19. Dowson, D.; Wragg, A. Maximum-entropy distributions having prescribed first and second moments. *IEEE Trans. Info. Theory* **1973**, *19*, 689-693.
20. Singh, V.P. *Entropy Based Parameter Estimation in Hydrology*; Kluwer Academic: Dordrecht, The Netherlands, 1998.
21. Souma, W. Universal structure of the personal income distribution. *Fractals* **2001**, *9*, 463-470.
22. Bouchaud, J.P.; Mezard, M. Wealth condensation in a simple model of economy. *Physica* **2000**, *A282*, 536-545.
23. Levy, M.; Solomon, S. Power laws are logarithmic Boltzmann laws. *Int. J. Mod. Phys. C* **1996**, *7*, 595-601.
24. Chatterjee, A.; Sinha, S.; Chakrabarti, B.K. Economic inequality: Is it natural? *Curr. Sci.* **2007**, *92*, 1383-1389.
25. Richmond, P.; Hutzler, S.; Coelho, R.; Repetowicz, P. A Review of Empirical Studies and Models of Income Distributions in Society. In *Econophysics and Sociophysics: Trends and Perspectives*; Chakrabarti, B.K., Chakraborti, A., Chatterjee, A., Eds.; Wiley-VCH: Berlin, Germany, 2006.

26. Gallegati, M.; Keen, S.; Lux, T.; Ormerod, P. Worrying trends in econophysics. *Physica A* **2006**, *370*, 1-6.
27. Willis, G.; Mimkes, J. Evidence for the independence of waged and unwaged income, evidence for Boltzmann distributions in waged income, and the outlines of a Coherent theory of income distribution; arXiv:cond-mat/0406694, 2004.
28. Sinha, S. Stochastic maps, wealth distribution in random asset exchange models and the marginal utility of relative wealth. *Phys. Scr. T* **2003**, *106*, 59-64.
29. Chatterjee, A.; Chakrabarti, B.K. Kinetic market models with single commodity having price fluctuations. *Eur. Phys. J. B* **2006**, *54*, 399-404.
30. Hayes, B. Follow the money. *Am. Sci.* **2002**, *90*, 400-405.
31. Hogan, J. Why it is hard to share the wealth. *New Sci.* **2005**, *2490*, 6-8.
32. Ball, P. Econophysics: Culture crash. *Nature* **2006**, *441*, 686.
33. Econophysicists matter. *Nature* **2006**, *441*, 667.
34. Lazear, E.P.; Shaw, K.L. Personnel economics: the economist's view of human resources. *J. Econ. Perspec.* **2007**, *21*, 91-114
35. Jones, K. C.E.O. Compensation: the pay at the top. *New York Times*, 4th April, 2009.
36. Etter, L. Hot topic: are ceos worth their weight in gold? *The Wall Street Journal*, 21st January, 2006.
37. Venkatasubramanian, V. A theory of design of complex teleological systems: unifying the darwinian and boltzmannian perspectives. *Complexity* **2007**, *12*, 14-21.
38. Jaynes, E.T. Where do We Go from Here? In *Maximum Entropy and Bayesian Methods in Inverse Problems*; Smith, C.R., Grandy, W.T., Eds.; Reidel Publishing: Boston, MA, USA, 1985.
39. Georgescu-Roegen, N. *The Entropy Law and the Economic Process*; Harvard University Press: Cambridge, MA, USA, 1971.
40. Samuelson, P.A. Gibbs in Economics. In *Proceedings of the Gibbs Symposium, American Mathematical Society*; American Mathematical Society: Providence, RI, USA, 1989.
41. Baumgartner, S. Thermodynamic Models: Rationale, Concepts, and Caveats. In *Modeling in Ecological Economics*, Proops, J., Safanov, P., Eds.; Edward Elgar Publishing: Northampton, MA, USA, 2005.
42. Rawls, J. *Justice as Fairness: A Restatement*; Belknap Press of The Harvard University Press: Cambridge, MA, USA, 2001.
43. Nozick, R. *Anarchy, State, and Utopia*; Blackwell: Oxford, UK, 1974.